\documentclass[fleqn,usenatbib]{mnras}
\usepackage{newtxtext,newtxmath}
\usepackage{graphicx}
\usepackage{newtxtext,newtxmath}
\usepackage[T1]{fontenc}
\usepackage{ae,aecompl}
\usepackage{amsmath}	
\usepackage{amssymb}	
\usepackage{xcolor}
\usepackage{mathtools}
\usepackage{hyperref}
\title[A stellar relic  filament]{A stellar relic  filament in the Orion star forming region\thanks{Based on observations done at Paranal under Prog. ID. 098.C-0850(A).}}

\author[Tereza Jerabkova et al.]{
Tereza Jerabkova$^{1,2,3}$\thanks{E-mail: tjerabkova@gmail.com},
Henri M.J. Boffin$^{1}$,
Giacomo Beccari$^{1}$,
and Richard I. Anderson$^{1,4}$
\\
$^{1}$European Southern Observatory, Karl-Schwarzschild-Stra\ss{}e 2, 85748 Garching bei M\"{u}nchen, Germany \\
$^{2}$Helmholtz Institut f\"{u}r Strahlen und Kernphysik, Universit\"{a}t Bonn, Nussallee 14-16, 53115 Bonn, Germany \\
$^{3}$Astronomical Institute, Charles University in Prague, V Hole\v{s}ovi\v{c}k\'ach 2, CZ-180 00 Praha 8, Czech Republic \\
$^{4}$D\'epartement d'Astronomie, Universit\'e de Gen\`eve, 51 Ch des Maillettes, 1290 Sauverny, Switzerland \\
}
\date{Received 6 June 2019; accepted 19 August 2019}
\pubyear{2019}

\begin{document}
\label{firstpage}
\pagerange{\pageref{firstpage}--\pageref{lastpage}}
\maketitle

\begin{abstract}
We report the discovery of the oldest stellar substructure in the Orion star forming region (OSFR), the \emph{Orion relic filament}. The relic filament is physically associated with the OSFR as demonstrated by Gaia DR2 photometry and astrometry, as well as targeted radial velocity follow-up observations of a bright sub-sample of proper-motion selected candidate members. Gaia DR2 parallaxes place the Orion relic filament in the more distant part of the OSFR, $\approx 430$~pc from the Sun. 
Given its age, velocity dispersion, spatial extent, and shape, it is not possible for the Orion relic filament to have formed as a single stellar cluster, even taking into account residual gas expulsion. The relic filament is also too young to be a tidal stream, since Galactic tides act on much longer time scales of order 100~Myr.
It therefore appears likely that the structure formed from a molecular cloud filament similar to Orion A in the OSFR and retained its morphology despite decoupling from its natal gas. Hence, the Orion relic filament bears witness to the short-lived evolutionary phase between gas removal and dispersion due to shears and tides, and provides crucial  new insights into how stars are formed in molecular clouds. 
\end{abstract}
\begin{keywords}
Stars: formation -- Stars: pre-main sequence -- Open clusters and associations: individual: ONC
\end{keywords}
%
\section{Introduction}
The Orion star forming region (hereafter OSFR) is the closest active star forming site that is producing massive stars. 
Together with other two prominent OB associations, Sco-Cen and Per OB2, it lies on the Gould 
Belt of young stars \citep{Lesh1968,deZeeuw1999}, that 
feature a spatially complex structure \citep{Zari2018}. The Gould Belt  of young stars spatially coincides with the structure known as the Lindblad Ring of HI \citep{Lindblad1973} --  
the Sun lying at a galactocentric radius within the Lindblad Ring.

The OSFR is known to have been forming stars over the last 12 Myr producing in total around $10^4\,M_{\odot}$ in several clusters and sub-groups confined to approximately 100--200 $\mathrm{pc}^3$. 
It is one of the best-studied star forming region. 
\cite{Bally2008_book} subdivides the OSFR into several groups based on sky distribution, distances, and ages: group 1a (8-12 Myr, 350 pc), 1b (3-6 Myr, 400 pc), 1c (2-6 Myr, 400 pc), 1d ($<2$~Myr, 420 pc) and $\lambda$ Ori ($<5$~Myr). 
The OSRF hosts an "integral-shaped filament" in the northern portion of the Orion A molecular cloud 
\citep{Bally1987}, within which the young Orion Nebula Cluster is forming  \citep{Hillendbrand1997, Beccari2017}. It is the largest molecular cloud/filament in the local neighbourhood \citep{Gorssschedl2018}.

Using Gaia \citep{2016A&A...595A...1G} DR2 photometry and astrometry, OmegaCAM photometry and spectroscopy with Hermes on the Flemish 1.2-m Mercator telescope we report here the serendipitous discovery 
of an $\approx$ 17 Myr old, cigar-like  ($\approx 90\,$pc long) structure that is clustered and distinct in proper motions and parallaxes. 
This newly found structure is the oldest stellar population linked to the ORSF and seems to represent the first case of a known stellar relic filament.

\section{Observations and datasets}
\label{Sec:Data}
We used the python Astroquery package 
\citep{gaia_py} to retrieve the Gaia DR2 data \citep{GAIA_DR2} and \citep{2018AA616A2L,2018AA616A3R,Evans2018} from the Gaia science archive\footnote{http://gea.esac.esa.int/archive/}. 
We downloaded all the objects detected by Gaia that are within a radius of fifteen degrees from the ONC (R.A.$ \approx 83.75 \, \mathrm{deg.}$, Dec $ \approx -5.48 \, \mathrm{deg.}$ -- 1 deg on sky corresponds to $\approx$ 7 pc at the ONC distance of 400 pc) without any additional filtering. Hereafter, we will refer to this catalogue as the Gaia or the 15-degree catalogue.  The size of the region described by the catalogue was made arbitrary, but is sufficiently large for the purpose of this study.

In this work we also used the photometric catalogue obtained via a set of deep multi-band images acquired with OmegaCAM (098.C-0850(A), PI: Beccari), a 1 deg$^2$ camera~\citep{kui11} attached to the 2.6-m VLT Survey Telescope (VST) at ESO's Paranal Observatory. The catalogue was used in \cite{Jerabkova2019} to study the stellar population in a $3\times3$ degree area around the centre of the ONC in the $r$ and $i$ filters. While the details of the data-reduction procedure are described in \cite{Jerabkova2019}, here we recap that the data were reduced and calibrated by the Cambridge Astronomy Survey Unit\footnote{http://casu.ast.cam.ac.uk/vstsp/}, while a large number of stars in common with the AAVSO Photometric All-Sky Survey (APASS) were used to correct for any residuals in the photometric calibration. The OmegaCAM catalogue includes 93,846 objects homogeneously sampled in the $r$ and $i$ bands down to $r \approx 21-22 \, \mathrm{mag}$ (ABmag) over a $3\times3~\mathrm{deg^2}$ area around the cluster's centre.

We used the $C^3$ tool  \citep{Riccio2017} in order to identify the stars in common between the OmegaCAM and the Gaia catalogue. $C^3$ is a command-line open-source Python script that, among several other options, can cross-match two catalogues based on the sky positions of the sources. 
We found 84,022 targets in common between the Gaia and the OmegaCAM data. The majority of objects that are present in the OmegaCAM catalogue, but not in Gaia, are typically faint ($r\gtrapprox 21$) and blue ($r-i \lessapprox 1.2$). 

We use Gaia DR2 parallaxes to separate members of the Orion star forming complex from fore- and background contaminators. Thus we apply a parallax selection criterion in the form
\begin{equation}\label{eq:C1}
2.0 - 3\sigma_{\varpi} \leq \varpi \leq 3.0 + 3\sigma_{\varpi} \,  \texttt{and} \, \sigma_{\varpi}/\varpi \leq 0.1 \, ,
\end{equation}
where $\sigma_{\varpi}$ is the uncertainty on the parallax as given in Gaia DR2. 
That is, we impose that the Orion candidate members are confined between distances of 333-500 pc from the Sun \citep[e.g.][]{Zari2019}.
 In addition, only the targets having a relative parallax error smaller than 
10\% are considered, to ensure a good quality of the astrometric solution~\citep[see also][]{GAIA_DR2}. Hereafter, we will refer to this catalogue as the ``3x3 deg catalogue''.

 We convert the observed angular proper motions (mas/yr) to {\bf tangential velocities} (km/s) by computing $v_{\mathrm{R.A.,Dec}}[\mathrm{km/s}]=\mu_{\mathrm{R.A.,Dec}}[\mathrm{mas/yr}]\cdot 4.74/\varpi[\mathrm{mas}]$. {\bf Such} velocities are better suited to search for kinematically clustered structures in the OSFR than observed proper motions, since the spatial extent of the OSFR is a significant fraction of the distance to it ($\Delta d/d \approx 0.4$). Thus, converting to  velocities resolves the distance degeneracy affecting observed proper motions in (mas/yr).

Since for the 15-degree catalogue we rely only on the Gaia
DR2 photometry, we also impose a photometric quality filter, using
the photometric excess criterion
\citep{2018AA616A2L,Evans2018,Arenou2018} defined as  
\begin{eqnarray}
1.0 + 0.015(G_{BP} - G_{RP})^2 < E < 1.3 + 0.06(G_{BP}-G_{RP})^2 \,,
\end{eqnarray}
where $E$ is 
defined as the flux ratio in the three different Gaia passbands, 
\begin{equation}
E =\frac{I_{BP} + I_{RP}}{I_G}\, .
\end{equation}

It is known that the Gaia DR2 photometry suffers from significant systematic effects in regions characterized by severe stellar crowding. This can be mitigated by using the criterion above. For more details, we refer to \citet{Jerabkova2019}.

\subsection{Spectroscopic observations with Hermes \label{sec:hermes}}

In order to verify that the group of stars discovered in proper motion space (see below) is kinematically  bound, we derived the line-of-sight (radial) velocities.
In January 2019, we selected 13 candidate member stars for spectroscopic follow-up with the goal of determining the third, radial, velocity component.
The 13 objects observed were selected based on their clear  membership indicated by the {\it Gaia} DR2 astrometry and photometry, as well as having a $V$-band magnitude brighter than  $\approx 13.5$\,mag to be accessible via a 1m-class telescope. 

The spectroscopic observations were carried out between 1 and 11 March 2019 using the high-resolution optical fibre-fed Echelle spectrograph {\it Hermes} mounted on the Flemish 1.2m Mercator telescope located on the Roque de los Muchachos Observatory on La Palma, Canary Island, Spain \citep{2011A&A...526A..69R}. All observations were carried out using the high-resolution fibre (HRF), which yields a spectral resolving power of $R \sim 85,000$ and the highest throughput. We selected exposure times between 180 and 1800s, aiming for signal-to-noise (S/N) ratios of about 15 around $\lambda 5000-6000$~\AA. The raw spectra were reduced using the dedicated {\it Hermes} reduction pipeline that carries out standard processing steps such as flat-fielding, bias corrections, order extraction, and cosmic clipping. A summary of our observations is available in Tab.~\ref{tab:targets_hermes}. 

The radial velocities were measured by cross-correlation with synthetic spectra, which were constructed at the same spectral resolution as the observations. Three synthetic spectra were used, depending on the brightness (i.e. the spectral type) of the target considered: a hot one ($T_{\rm eff}=8000$~K, $\log g= 4.$), a solar model, and a cooler star ($T_{\rm eff}=5000$~K, $\log g= 4.5$). The models were created with the software\footnote{\url{http://www.appstate.edu/~grayro/spectrum/spectrum.html}} {\tt SPECTRUM v2.76} of R.O.~Gray. 

The measured radial velocities are listed in Tab.~\ref{tab:targets_hermes}.

\begin{figure*} 
 \scalebox{1.0}{\includegraphics{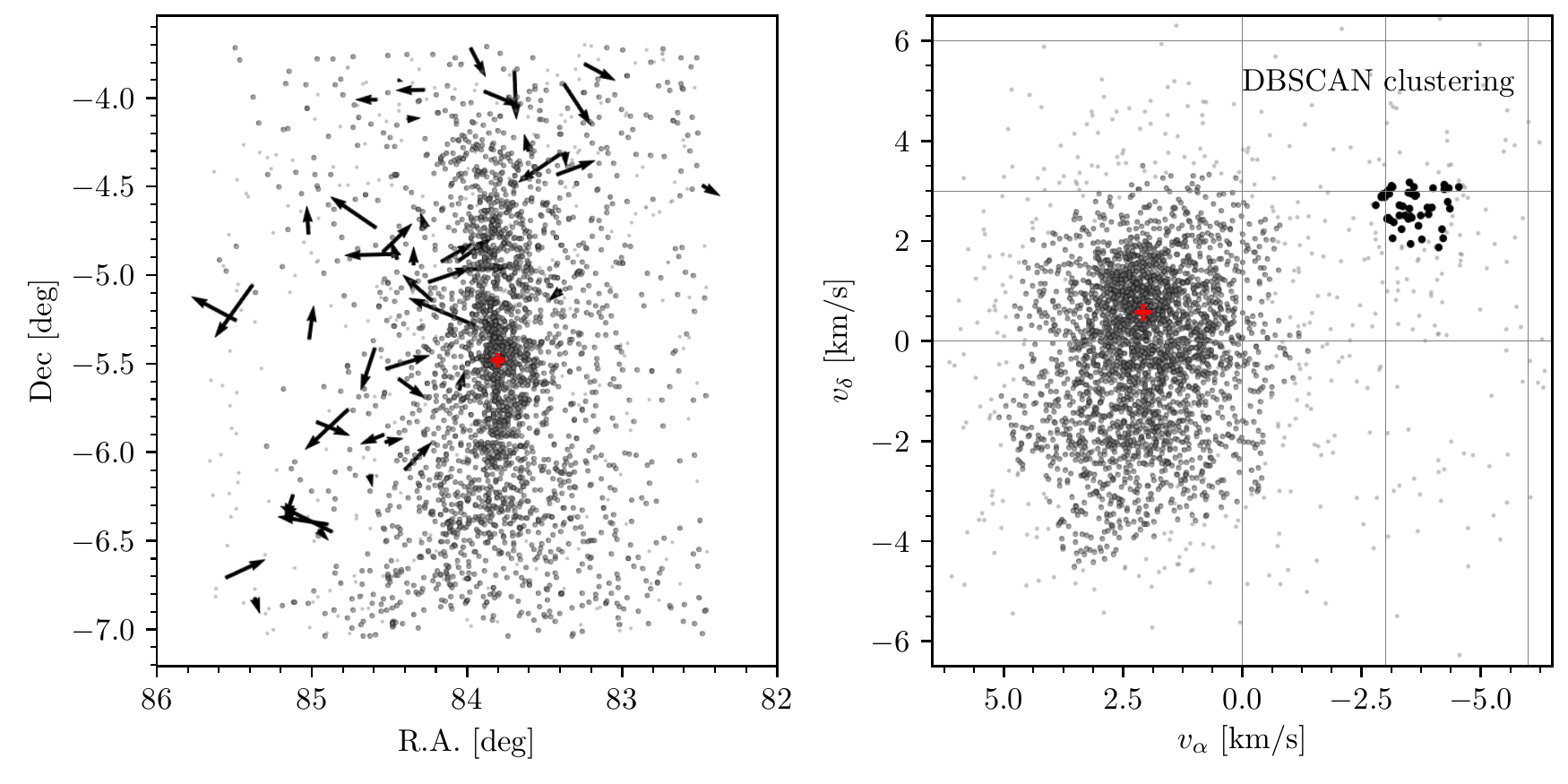}}
 \caption{
\textbf{Right panel:} 
 Distribution of proper motions of the targets in the initial catalogue after parallax selection.
 The faint gray points show all targets, while the black and dark gray points were identified by the DBSCAN clustering algorithm as two distinct groups. The red point represents the peak value of the distribution of proper motions of the Orion Nebula Cluster (ONC) \citep{Jerabkova2019}. The vertical gray lines are plotted for values of $v_{\alpha}$ of 0, -3 and -6 km/s. The horizontal gray lines are plotted for values $v_{\delta}$ of 0, 3 and 6 km/s.
 \textbf{Left panel:} The distribution of the targets on the sky with the same colour coding as in the right panel. The arrows indicate the proper motions of the black points after removing the mean value. 
The red cross indicates the position of the Orion Nebula Cluster for reference.}
\label{fig:blu_cl}
\end{figure*}

\section{Discovery of the Orion Relic Filament}

\subsection{A group/cluster in proper motions}
While investigating the three bursts of star formation in the Orion Nebula Cluster 
\citep[ONC; ][]{Jerabkova2019}, using the OmegaCAM catalogue mentioned in Sec.~\ref{Sec:Data}, we noticed in  proper motion space a distinct clump of stars  that is well separated from the ONC. The sky distribution (R.A., Dec) and 
the proper motion distribution ($v_{\alpha}$,$v_{\delta}$) are plotted in 
Fig.~\ref{fig:blu_cl} (left and right panel, respectively). We run the clustering algorithm using DBSCAN \citep{Ester96adensity-based} in a three-dimensional plane containing the two proper motions on sky and the parallax, and are able to identify and thus confirm the presence of two clear structures -- the larger structure containing the Orion A stellar population (in gray) and a distinct clump (black) at $v_{\alpha}\approx -3.8$ km/s and $v_{\delta}\approx 2.5$ km/s, as seen in Fig.~\ref{fig:blu_cl}.

\begin{figure*}
 \scalebox{0.9}{\includegraphics{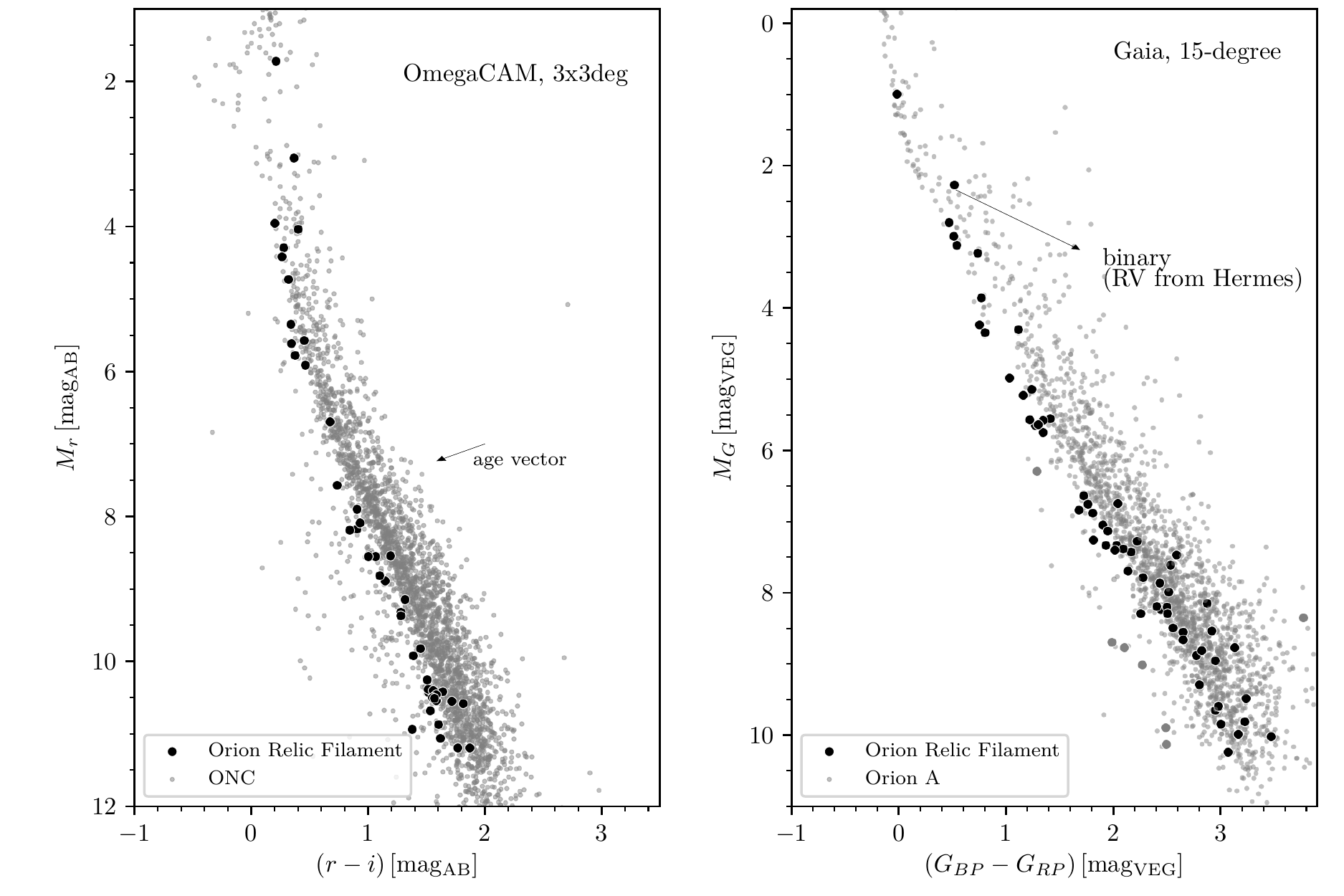}}
 \caption{\textbf{Left panel:} 
 Colour-absolute magnitude diagramme (CMD) of the $3 \times 3\deg^2$ catalogue using OmegaCAM $r$ and $i$ filters. Gaia parallaxes were used to compute the absolute magnitudes.
 The gray and the black points show the stars in the clusters identified in proper motions and parallax by the DBSCAN algorithm and shown in Fig.~\ref{fig:blu_cl}. 
 \textbf{Right panel:} The CMD for the Gaia 15-degree catalogue using the Gaia photometric data. In gray, we see all the already known, stars identified in the catalogue with DBSCAN using on-the-sky distributions (see also Fig.~\ref{fig:Orion_radec} and Fig.~\ref{fig:Orion_onsky}). The black points correspond to the 
 proper motion-selected group members in the the OmegaCAM catalogue -- the Orion relic filament.
}
\label{fig:blu_CMD}
\end{figure*}

The objects belonging to the distinct clump form an elongated structure on the sky and do not overlap with Orion A. The parallax distribution peaks at a value corresponding to a distance of 430 pc ($\pm 20$ pc), placing the objects from the clump further away from us than the ONC.
In addition, the structure is co-eval in the colour-magnitude diagram (CMD), being older than the rest of the population in  Orion A, as one can see in the left panel of Fig.~\ref{fig:blu_CMD}.
The cluster in proper motions was also seen by \cite{Zari2019} for Orion A region where co-evality was discussed (in the paper noted as group F and see also their group B8) and by \cite{Kounkel2018}, but without further discussion or analysis.

\subsection{Extent of the discovered feature}

By inspecting the distribution of the stars on sky as shown on the left panel of Fig.~\ref{fig:blu_cl} we cannot exclude that 
the newly discovered population extends beyond the area covered by the 3x3 deg catalogue. We hence decided to use the 15-degree catalogue with the aim to recover the structure on a wider area. 
It should be noted that the stellar populations in the OSFR is complex. In particular, 
if we consider a larger area than 3x3 deg, then in proper motion space, other populations from the region tend to blend in with the structure discovered in the 3x3 deg catalogue, making it much less prominent.
In light of the OSFR's complexity and angular size, we apply two independent, proper-motion-based methods (S1 and S2, described below) to the 15-degree catalogue and check whether the clump can be successfully recovered.

\begin{figure*}
\scalebox{0.9}{\includegraphics{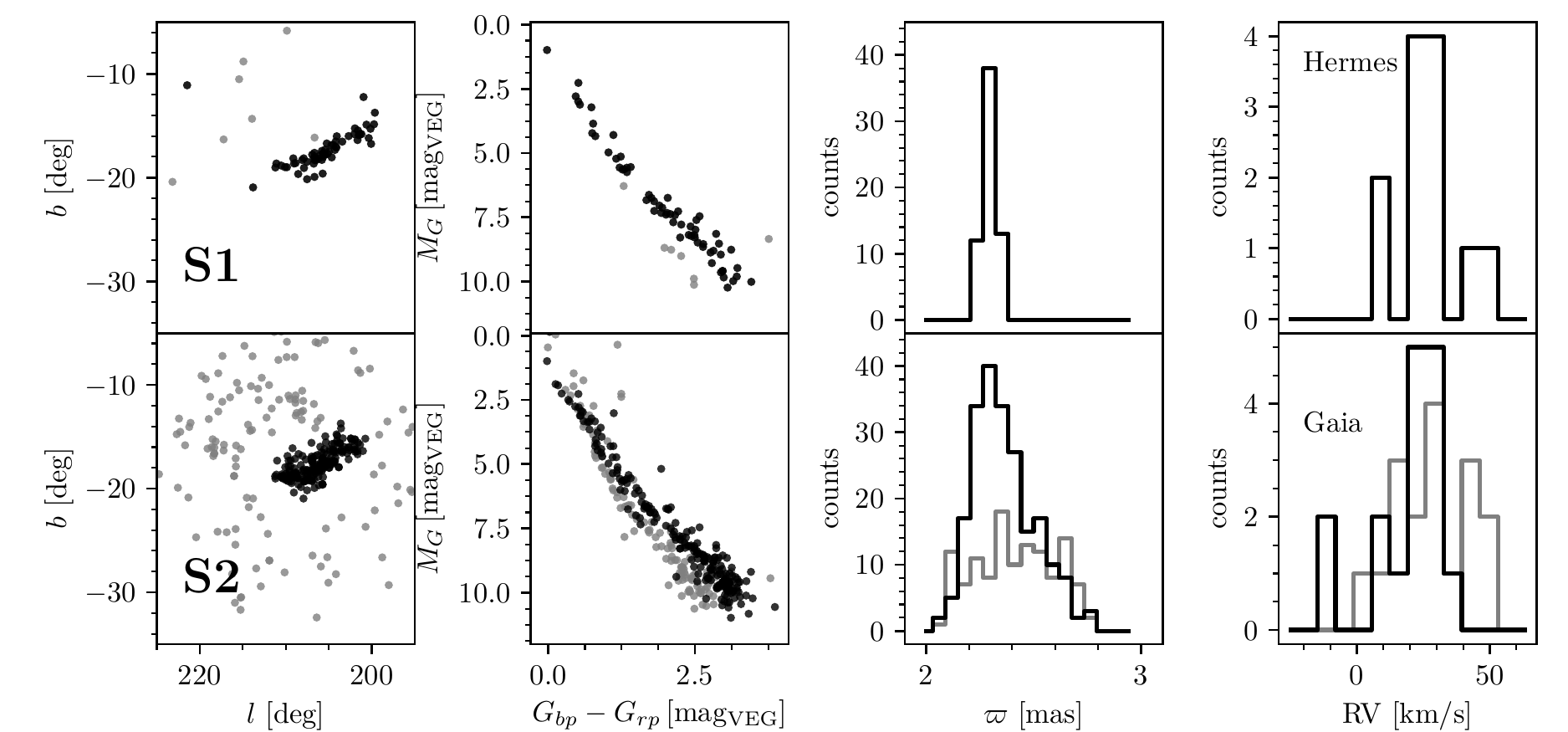}}
 \caption{Distribution in the Galaxy (uttermost left) and in the colour-magnitude diagramme (second column) for the selection using either S1 (top) or S2 (bottom) -- see text. The associated parallax distribution is show in the third column and the Hermes(top)/ Gaia(bottom) radial velocity distribution (right). 
The black points are those identified as members depending on the selection criterion, the gray points show non-members. We note that the CMD on the top panel shows the same data sample as right panel of Fig.~\ref{fig:blu_CMD}, the CMD on the bottom panel contains fore-/back-round contaminators. 
The gray histograms for the bottom panels show distributions for targets that are not identified as members.}
\label{fig:filament}
\end{figure*}

\textbf{S1)} We select only the stars having parallaxes in the range $2.16<\varpi<2.48$ as defined by the stars in the clump in proper motions found in the 3x3 deg catalogue. 
This approach is lowering the contamination from the population of the OSFR at the distance where the objects belonging to the clump is expected.
We then run DBSCAN on this sub-catalogue in the 3D space spanned by proper motions and parallax. The stars belonging to the new clump are successfully isolated. Their distribution on the sky and their CMD are shown in Fig.~\ref{fig:filament} (black points on the panels on the top left and second top row, respectively). Since the method S{\bf1} relies only on the clustering algorithm DBSCAN, we do additional checks of the physical properties not used in the clustering -- radial velocities and the CMD. In addition, we introduce the method S2 that is clustering-independent.

\textbf{S2)} We consider all targets in the 15-degree catalogue that have proper motions in the circular region corresponding to the clump as derived from the 3x3 deg catalogue (that is, a radius of 1.0 km/s centred around $v_{\alpha}$=-3.8  km/s,$v_{\delta}$=2.5 km/s). The bottom panels of Fig.~\ref{fig:filament} show the target locations in Galactic coordinates as well as the CMD, parallax distribution, and any available radial velocity data of these targets in the 15-degree Gaia catalogue. By way of construction, the method S2 will lead to a much larger contamination from fore- and background stars, as clearly seen in the figure. These can be separated when using the CMD.

Using  method S1 we can clearly isolate a clump of stars in proper motion as previously done in the 3x3 deg catalogue. The clump identified in the 15-degree catalogue is located at the same position as the original clump in the 3x3 deg catalogue, but is more extended, therefore
bringing the values of proper motions of the members of the clump closer to those of the Orion A members plotted in Fig.~\ref{fig:blu_cl}.
Using this selection we can confirm the presence of a coeval population of stars which extends well beyond the 3-degree area studied before. 
We note the recovered population's elongated, filamentary morphology in Galactic coordinates, whereas it is clearly clustered in proper motion space.  The S1 method requires a significant cut in parallaxes in order to discern in the 15-degree Gaia data set the clump in proper motions that was detected in the 3x3 deg catalogue. 
Due to  still large uncertainties of Gaia DR2 parallaxes (10\% error is $\approx$ 40pc at the distance of the OSFR),  the S1 method is not able to recover all the targets belonging to the structure by using S1 but only the subset of likely members.

 Method S2 allows us to explore a larger region in parallax values. This method consistently reveals the presence of an over-density of stars on sky at the same position and with the same extent as {for objects selected by method} S1. The distribution on the CMD of the stars populating the elongated over-density (black points on the first and second lower row plot of Fig.~\ref{fig:filament}) confirms that these stars are coeval. They are, however, younger than the sparse component (in gray on the same plots). Furthermore, the parallax distribution of such stars is statistically consistent with the distribution  imposed via the S1 method (see the black histograms on the plots of the third column). The method S2 clearly introduces more contamination into the discovered structure from the field stars than method S1. On the other hand, the total extent of the structure is the same as for the method S1. That is, the extent seen can thus be considered real and the size of the catalog 15deg sufficient.

 At the request of an anonymous referee we performed two additional tests of the robustness of method S1. The main principle of both is to randomly reshuffle proper motions in R.A and Dec independently while keeping other catalogue parameters intact and test what structures can be recovered by the clustering algorithm. Test one  uses the same data set as in S1 and test two applies a cut in $l$ and $b$ to select the area around the discovered structure only. For test one, we reshuffled the proper motion values several times and were not able to find similar structures to the one we have detected. We note that even when a larger structure was found in the reshuffled proper motions, it never had a co-eval distribution in the CMD nor similar values of RVs. This means that the clustering algorithm is robust and the discovered structure is not a density fluctuation in proper motion space.
 For test two, we applied a cut in $l$ and $b$ as shown in the middle left panel of Fig.~\ref{fig:lb_cut}. 
 The plot in the middle row right panel of Fig.~\ref{fig:lb_cut} shows that the cut in $l$ and $b$ described above clearly lowered the overall contamination, thereby increasing the apparent visibility of the clump in proper motions. However, the contamination from the OSFR is still too high to allow the DBSCAN algorithm to recover the discovered clump of stars. Thus, we further restricted the catalogue in parallax as shown in the bottom row of Fig.~\ref{fig:lb_cut}.  We were able at this point to fully recover the clump in proper motions using DBSCAN.
 
\begin{table*}
    \centering
    \begin{tabular}{c|c|c|c|c|c|c|c|l}
         ID & R.A. & Dec. & $G$ & $RV_{\mathrm{Gaia}}$ & $RV_{\mathrm{Hermes}}$  & Epoch  & Exp. time & comment  \\ 
         & & & [mag] & [km/s] 
         & [km/s] 
         & (BJD-245854) & [s] & \\
         \hline
3216997407412184576  & 82.16820 & -2.73367 &  9.14  &  --  &  +13.5 $\pm$ 2.6  & 4.33881 & 300 &  binary?\\ 
3016216856077986816  & 85.36432 & -6.81766 &  10.46 &  --  &  $-55.7 \pm 4.1$  & 4.3436 & 180 &  binary! \\ 
  &  &  &   &   &  $-41.4 \pm 4.1$  & 9.35148 & 500 &  \\ 
3320525395058674176  & 86.46964 & 4.80873 &  10.93 &  --  &  +23.3 $\pm$ 2.7  & 5.34945 & 300&  \\ 
3223717107084373504  & 83.51508 & 2.63526 &  11.18 &  +34.16 $\pm$ 1.50  &  +30.4 $\pm$ 2.0 & 5.35456 & 300&  \\ 
3223574204932246016  & 83.16921 & 2.19963 &  11.30 &  --  &  +29.2 $\pm$ 1.8  & 4.34802 & 300&  \\ 
3215983485892366848  & 83.47961 & -3.15763 &  11.39 &  --  &  +17.5 $\pm$ 1.2  & 6.3657 & 600& binary? \\ 
3010622094238648320  & 84.39491 & -10.11905 &  11.97 &  +50.94 $\pm$ 2.47  &  +50.1 $\pm$ 1.7  & 6.37477 & 700&  binary?\\ 
3236001709982983680  & 82.10707 & 3.33701 &  12.39 &  --  &  +45.4 $\pm$ 1.7  & 5.3586 8 & 180& \\ 
3217571696079288320  & 83.59811 & -1.36120 &  12.42 &  +25.15 $\pm$ 9.91  &  +28.9 $\pm$ 1.4 & 6.38637 & 1000&   \\ 
 &  &  &   &   &   +28.1 $\pm$ 1.8 & 11.3871 & 1800&  \\ 
3000057956459742720  & 96.68656 & -12.56129 &  12.51 &  +1.31 $\pm$ 3.79  &  +3.0 $\pm$ 1.7  & 5.36791  & 900&  binary?\\ 
3216096632510194176  & 84.04834 & -2.74891 &  13.15 &  --  &  +28.3 $\pm$  2.0  & 6.40058 & 1200&  \\
3216974386386913280  & 83.57078 & -1.93627 &  13.31 &  --  &  +27.4 $\pm$ 1.5  & 9.40488 & 1800&  \\
3223703603706959360  & 83.51694 & 2.54048 &  13.39 &  --  &  +31.3 $\pm$ 1.5  & 4.35853 & 1200& \\
    \end{tabular}
    \caption{List of the brightest members of the Orion relic filament followed up by spectroscopic measurements using the Hermes spectrograph on the Flemish 1.2m Mercator telescope. The targets are shown as white dots in Fig.~\ref{fig:Orion_radec}. When removing the targets with large RV variations, i.e. those that are likely close binaries, the velocity dispersion of our measurements is $\approx$ 4 km/s confirming the physical existence of the Orion relic filament.}
    \label{tab:targets_hermes}
\end{table*}

 We then randomly re-shuffled the proper motion values, as done in  test number one, and used DBSCAN to identify potential structures in proper motion space.  We did not find a single case in which, after re-shuffling of proper motions, the identified structures by DBSCAN would be coeval or share similar radial velocity values. This further confirms the statistical strength of our results and our conclusions.

In summary, we successfully recover the new kinematic structure in the wider, 15-degree catalogue.
We used two independent methods to recover the structure. For each method, we ensured that additional data -- that were not used in the identification of the structure -- are consistent with the existence of a coeval (CMD), kinematically unique (RVs), physical structure at the same distance (parallax distribution).
On the sky, the identified stars resemble a 90-pc long filament, henceforth referred to as the  {\it Orion Relic Filament},  which is older than Orion A, as indicated by the CMD.

\subsection{Radial velocity measurements\label{RVs}}

The RV measurements based on the Hermes spectra (cf. Sec. \ref{sec:hermes}, Tab.~\ref{tab:targets_hermes}) agree to within the errors with Gaia mean velocities, where these are available from DR2.
For two stars, we have had two measurements, separated by a few days. In one case,  Gaia DR2 3016216856077986816, we have clearly identified a spectroscopic binary, as over a period of 5 days, the radial velocity varied by 14 kms$^{-1}$. The position in the colour-magnitude diagram of this star also confirms its binary nature. This star would have a mass of about 2 M$_\odot$ and such stars are known to be often in binary systems. Given the mean radial velocity of the relic filament (see below), the minimum semi-amplitude of the radial velocity curve is at least 30 kms$^{-1}$, which implies a period above 10 days, but below $\approx 150$ days. For the second star for which we had two epochs, Gaia DR2 3217571696079288320, we did not detect any significant radial velocity change. Moreover, the value we obtain is compatible with that determined by Gaia, and with the mean value for the velocity of the relic filament. Hence, there is no indication that this star is a spectroscopic binary. 

We show on the right-hand panels of Fig.~\ref{fig:filament} the distribution of radial velocities from Gaia (lower panel) and  Tab.~\ref{tab:targets_hermes} (upper panel). For the latter we exclude all outliers. It is immediately seen that the distribution of the radial velocities of the stars belonging to the filament and identified via S1 and S2 are consistent.

Using the Hermes radial velocities we derive a mean value of 26.1 $\pm$ 4.7 kms$^{-1}$, which is consistent with the peak value, 28 $\pm$ 2 kms$^{-1}$, of the Gaia RV distribution based on the selection method S2 (see right-most panels in Fig.~\ref{fig:filament}). The dispersion is formally larger than the one seen in the proper motion space, but we note that it corresponds to only about twice the errors on our measurements (limited by the relatively low S/N of our spectra), so that it is most likely overestimated. There are four additional stars that have measured radial velocities which are different by more than 3-$\sigma$ from the mean velocity. These are likely spectroscopic binaries or, alternatively, stars that have been ejected. 
We note that if we accept that these are binaries, we would have a fraction of binaries of 5/13= 38\%, which is not unusual, especially as we are biased to the brightest objects in the filament, which could be brighter because of the presence of a companion.  

\begin{figure}
 \scalebox{.9}{\includegraphics{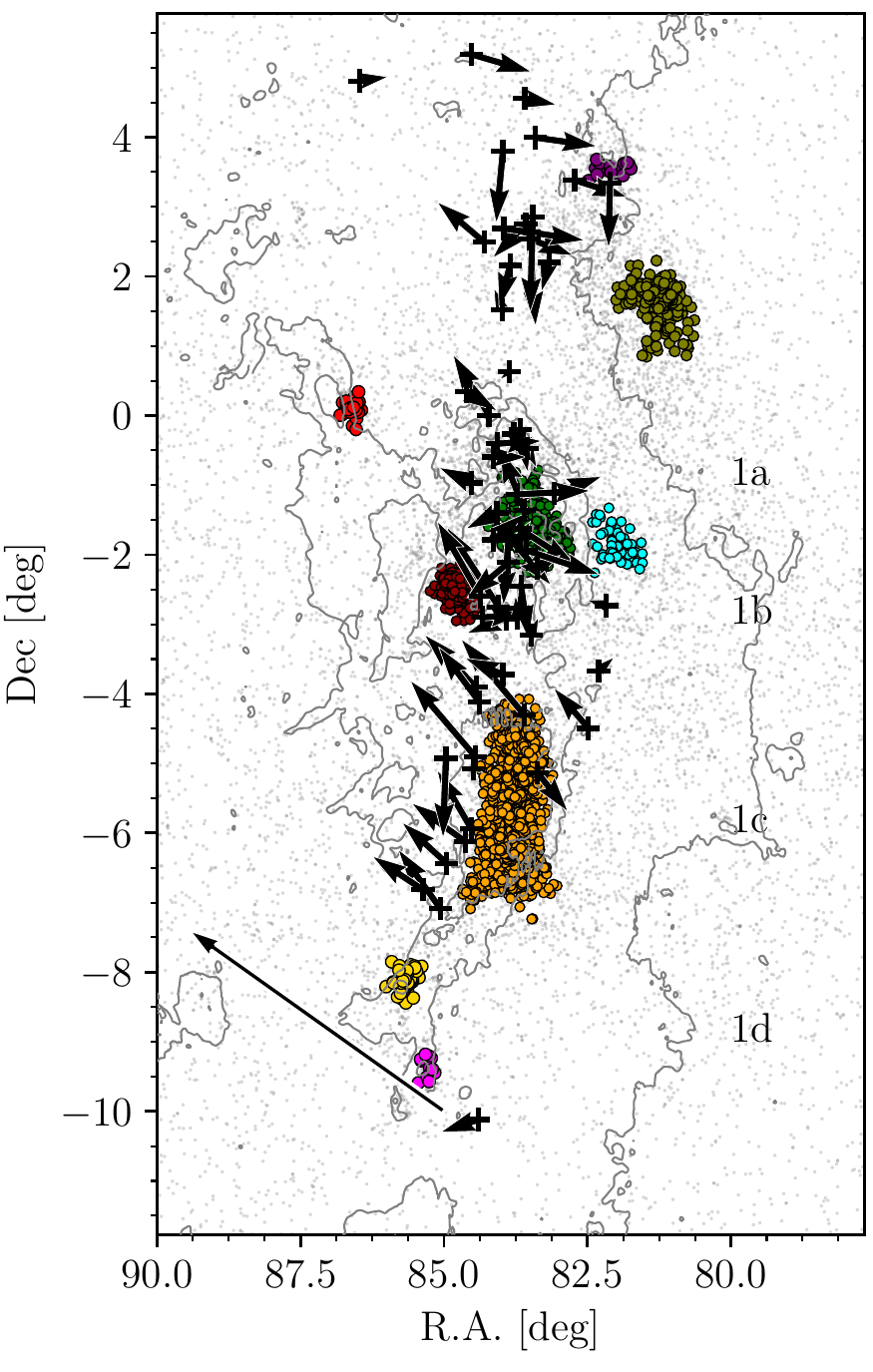}}
 \caption{The OSRF and the identified clusters in R.A. and Dec. All data points in the Gaia DR2 catalogue having parallaxes in the interval [2--3] mas are plotted as gray points. We used the DBSCAN algorithm to identify clustering in space (b,l, $\varpi$) -- the resulting core members of found groups are marked as colour points (see text for more details). The underling isocontours show IRAS 100 $\mu m$. The newly discovered old (17 Myr) structure is shown as black '+' symbols, while the white arrows show the proper motions in R.A. and Dec coordinates after removing the mean (indicated by the large black  arrow).
The division groups are listed in Tab.~\ref{tab:my_label}  and their names are placed on the right at the corresponding values of Dec.
 }
\label{fig:Orion_radec}
\end{figure}

\begin{figure*}
 \scalebox{.9}{\includegraphics{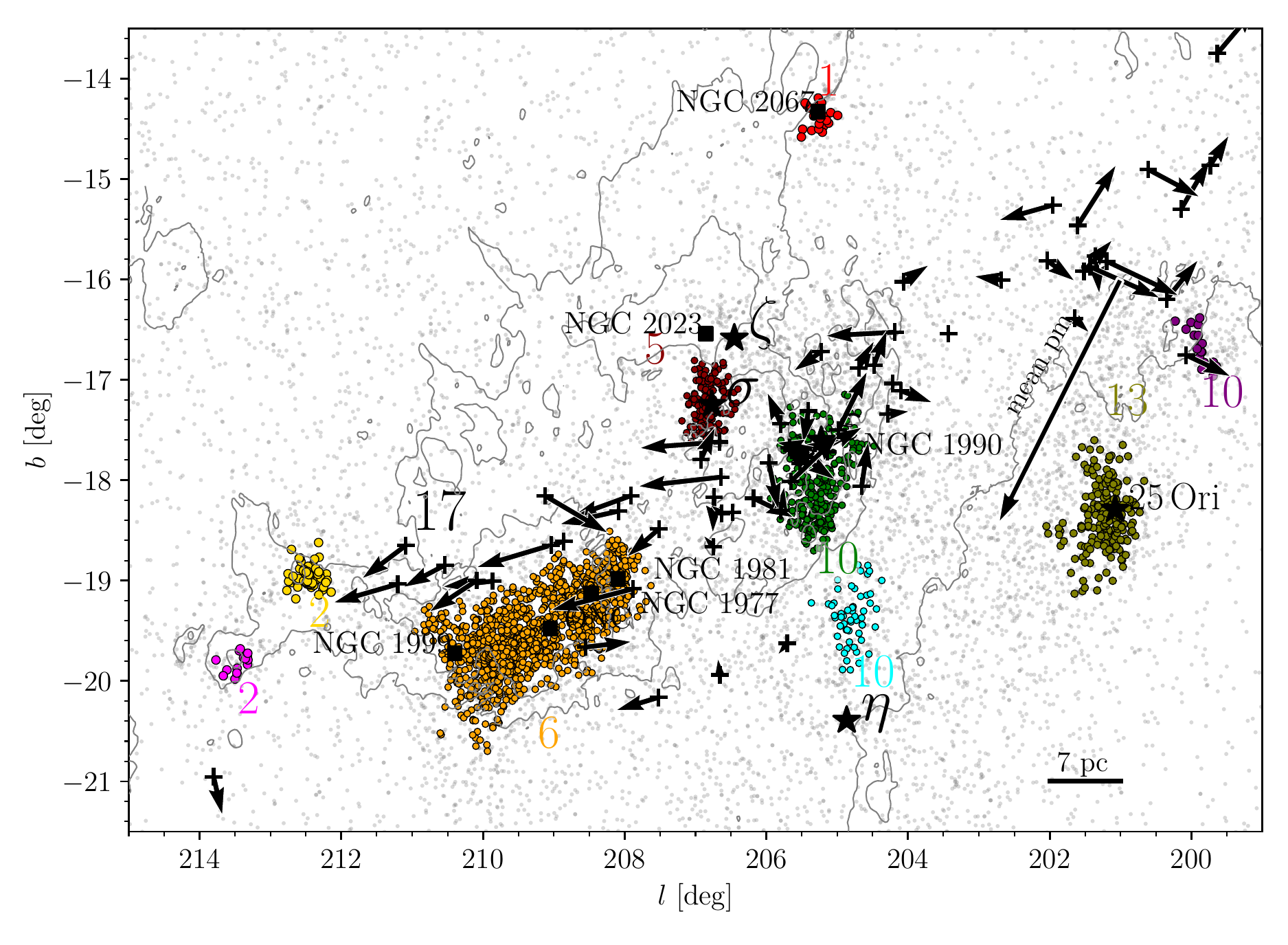}}
 \caption{The Orion region in galactic coordinates, $(l,b)$. All data points in the Gaia DR2 catalogue having parallaxes in the interval [2--3] mas are plotted as gray points. We used the DBSCAN  algorithm to identify clustering in space (b,l, $\varpi$) - the resulting core members of found groups are marked as colour points with estimated age based on isochrone fitting (see text for details) written next to them. The known star clusters are marked as squares and bright stars as star symbol. The underling isocontour shows IRAS 100 $\mu m$. The newly discovered old (17 Myr) structure is shown as black `+' symbol, while the white arrows show the proper motions in Galactic coordinates after removing the mean (indicated by the large black  arrow). 
}
\label{fig:Orion_onsky}
\end{figure*}

\subsection{Orion Relic Filament and Orion star forming region}

The next step is to put the discovered Orion relic filament into the context of the OSFR. For this purpose we investigate further the 15-degree catalogue. 
First, we run the clustering algorithm DBSCAN on the entire catalogue in the 3D parameter space [R.A., Dec, $\varpi$] to select substructures in the OSRF. 
We identified 9 groups or clusters that are clearly visible as over-densities of stars on the sky. We plot them in~Fig.~\ref{fig:Orion_radec}, and in Galactic coordinates, in Fig.~\ref{fig:Orion_onsky}, together with the known bright stars and star clusters. These groups are used as reference stellar populations of the OSFR. Note that even though a much more detailed analysis of the OSRF could be done \citep[for example, see][]{Kounkel2018}, we defer this to future work and here aim to  define the reference populations.  

\begin{figure*}
 \scalebox{0.9}{\includegraphics{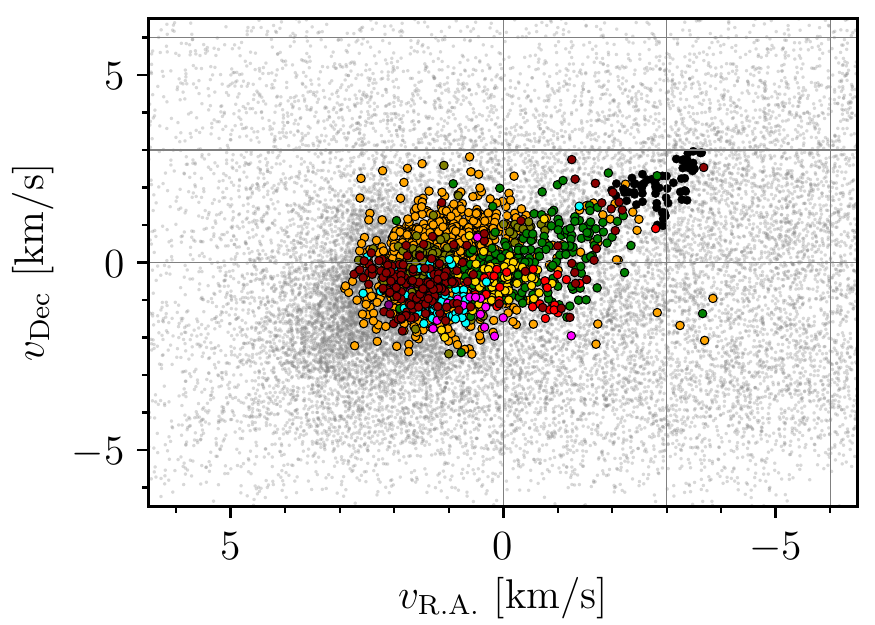}}
 \scalebox{0.9}{\includegraphics{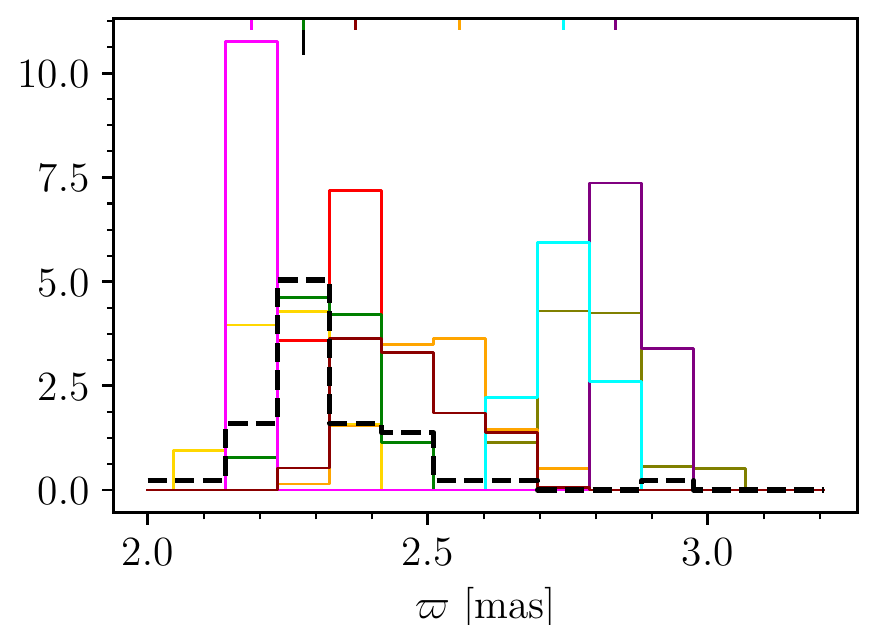}}
 \caption{
 \textbf{Left Panel:} The proper motions in R.A. and Dec coordinates). Different colours correspond to the different groups identified in Fig.~\ref{fig:Orion_onsky}, while the black points are members of the Orion relic filament. The vertical gray lines are plotted for values of $v_{\alpha}$ of 0, -3 and-6 km/s. The horizontal gray lines are plotted for $v_{\delta}$ values of  0, 3 and 6 km/s. \textbf{Right Panel:} The parallax distributions for the different groups identified in Fig.~\ref{fig:Orion_onsky}, with the black histogram showing the parallax distribution of the Orion relic filament. 
}
\label{fig:pm_par}
\end{figure*}

The extent of the Orion relic filament is similar to that of the OSFR as shown in~Fig.~\ref{fig:Orion_radec}  and Fig.~\ref{fig:Orion_onsky}. The typical proper motions of its stars are offset from the OSFR members (see the right panel of Fig.~\ref{fig:pm_par}). The distribution of parallaxes (Fig.~\ref{fig:pm_par}) clearly reveals that the Orion relic filament is located in the most distant part of the OSFR.

\begin{figure*}
 \scalebox{0.9}{\includegraphics{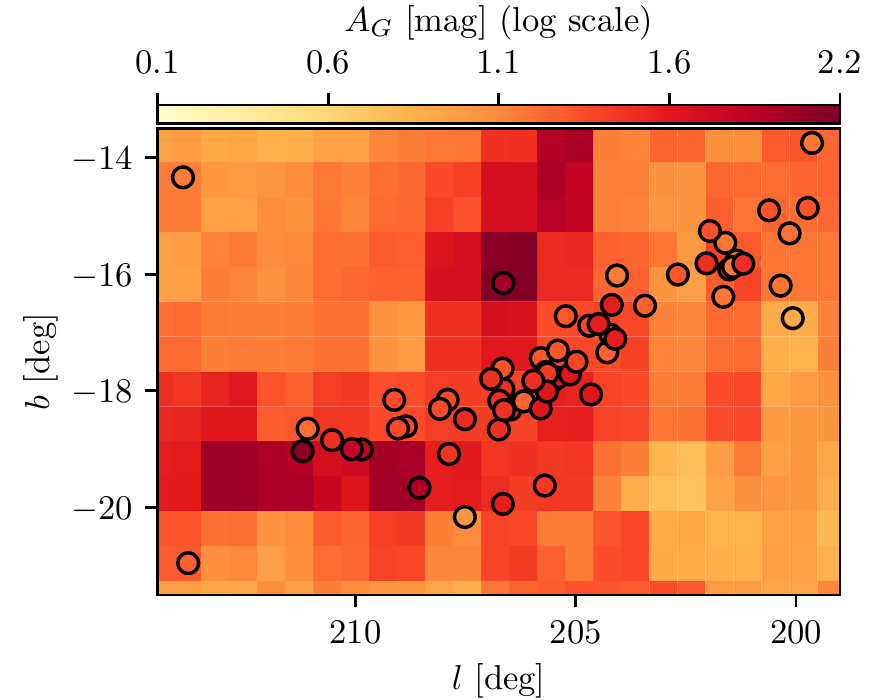}}
 \scalebox{0.9}{\includegraphics{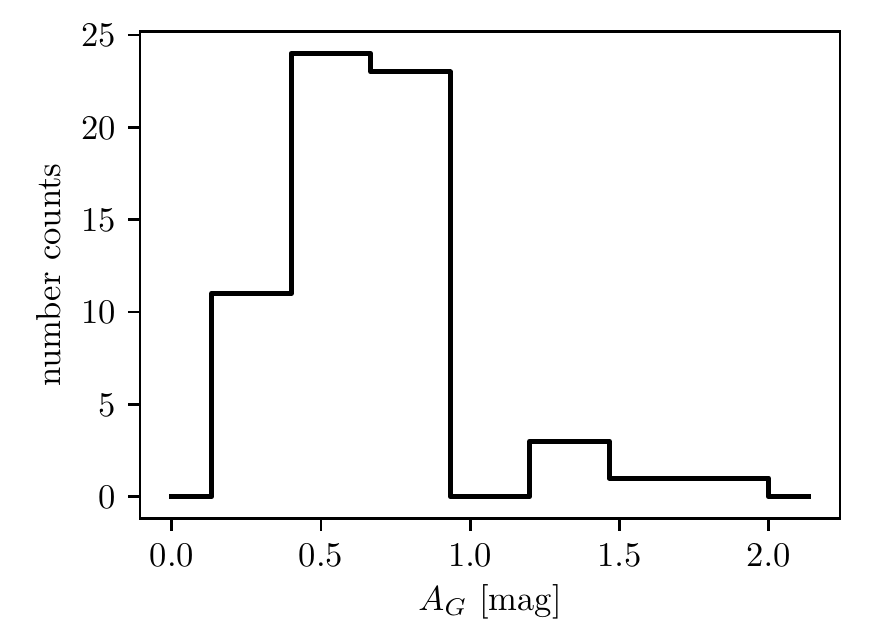}}
 \caption{
 \textbf{Left Panel:} Extinction map for the parallax cut $2.25 < \varpi < 2.40$, corresponding to the location of the Orion relic filament. The plotted points are the members of the Orion relic filament, with the grid boxes being colour-coded based on the computed values of $A_G$. 
  \textbf{Right Panel:} Distribution of $A_G$ values for the members of the Orion relic filament.
}
\label{fig:AG}
\end{figure*}

In order to further characterise the relic filament, we derived the age of each star by comparing its position on the CMD with stellar isochrones. Given the extent of the stellar structure and the likely presence of a foreground molecular cloud, some degree of differential extinction can be expected. Following the approach described by~\citet[][]{Zari2018} we use the interstellar extinction ($A_G$  and $E(G_{PB}-G_{RP})$) available for the brightest stars in the Gaia DR2 catalogue to build a 3D extinction map. In short, we first bin the data in $l, b$ and parallax space. For each cell we then calculate the mean and the standard deviation of $A_G$ and $E(G_{PB}-G_{RP})$ using the measurements available in the given region. We consider such values as characteristic values describing the amount of extinction in a given bin which is then used to de-redden the magnitudes of each star located in the same bin (Fig.~\ref{fig:AG}).
This allows us to put all stars of a given cluster in a colour-magnitude diagramme and determine their age by comparing with stellar isochrones. 
For this purpose we acquire a grid of solar metallicity PARSEC isochrones 
\citep[version 1.2S;][]{Bressan2012,Chen2014,Tang2014,Chen2015} with time steps of 0.1 Myr. 
 
\begin{figure}
 \scalebox{1.0}{\includegraphics{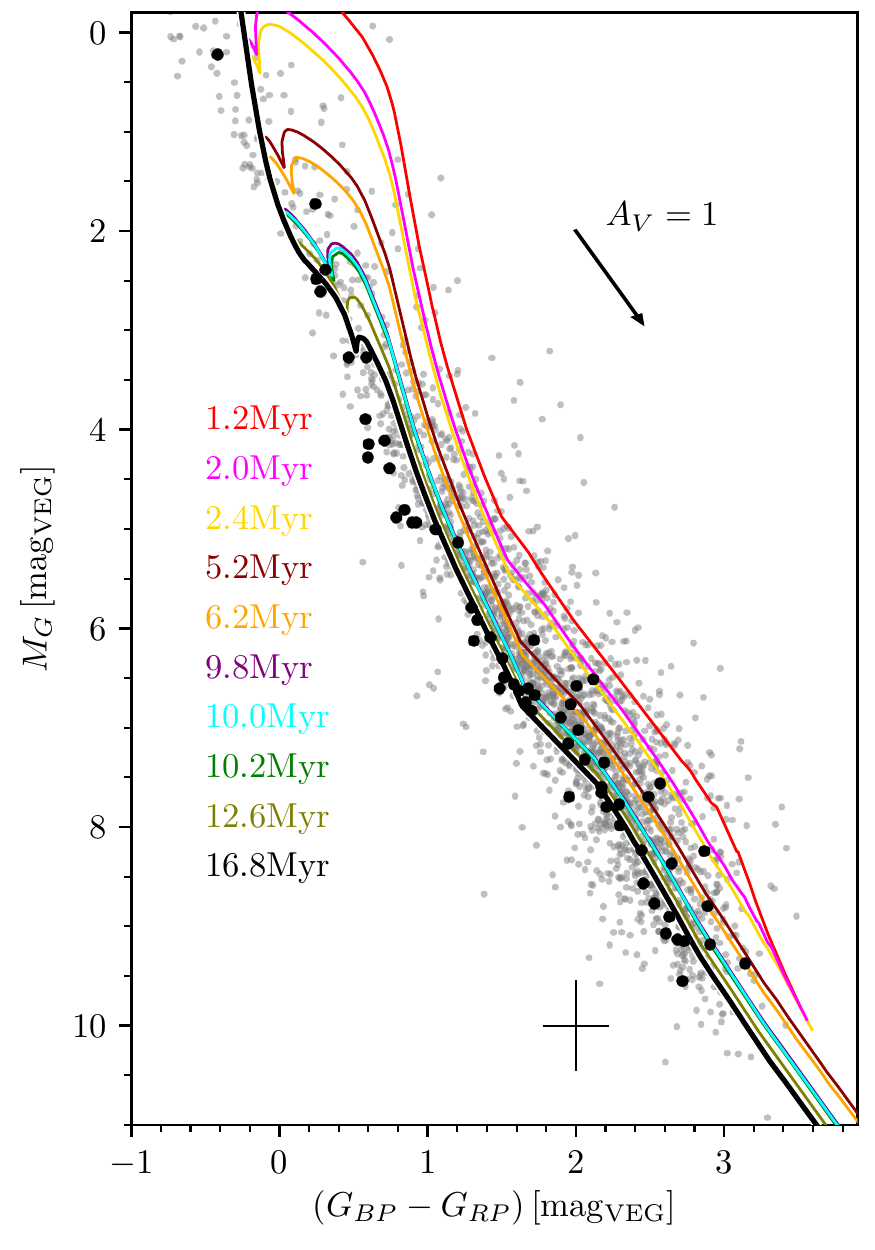}}
 \caption{
 The colour-absolute magnitude diagramme showing all the identified members of the OSFR in gray and the members of the Orion relic filament in black. The error bars for the black points combine the photometric uncertainties, the parallax uncertainties used to compute absolute magnitudes and the uncertainty given by the correction for the extinction: their representative value is only shown by the cross in the left bottom corner in order to increase the clarity of the plot. We do not show the error bars for the gray points.  The plotted isochrones correspond to the relic filament (black) and to the other identified structures in the OSFR. Tab.~\ref{tab:my_label} lists each age with the corresponding color, while the ages are also shown directly in this Figure. 
 }
\label{fig:CMD_age}
\end{figure}

 We run a least-square fitting routine to estimate the best age for each identified group in the OSFR and of the Orion relic filament. The errorbars on the best fitted age are estimated from the scatter of the points along the isochrones -- the age corresponding to a $1-\sigma$ offset from the mean is used. The final ages  and their uncertainties are quoted in Tab.~\ref{tab:my_label} and the best fitting isochrones are shown in Fig.~\ref{fig:CMD_age}. We recover the ages for the groups of the OSFR as quoted \cite{Bally2008_book}. The Orion relic filament is older than the rest of the OSFR, or at the very least, given the uncertainties, among the oldest populations present in the region. Thus this results suggests that the Orion relic filament was the first star formation event in the OSFR and was then followed by the rest of the star formation activity in this region. 

 To summarise, the Orion relic filament is the oldest kinematically distinct structure in the vicinity of the OSFR. It is several Myr older than the oldest previously known regions, with which it overlaps in ($l,b,\varpi)$ space (cf. Fig.~\ref{fig:Orion_onsky}).

\begin{table}
    \centering
    \begin{tabular}{l|l|c|r}
         Name & colour & mean($\varpi$ [mas]) & age [Myr]  \\
         \hline
  1d-1     & magenta & 2.3 $\pm$ 0.2 & 2.0 $\pm$ 0.6\\
  1d-2       & yellow & 2.3 $\pm$ 0.2  &  2.4 $\pm$ 0.6\\
  1c-1      & gold  &  2.6 $\pm$ 0.3 & 6.2 $\pm$ 0.5 \\
  1b       & dark-red & 2.5 $\pm$ 0.3 & 5.2 $\pm$ 1.5\\
  1a-1       & green & 2.4 $\pm$ 0.3 & 10.2 $\pm$ 0.6\\
  1a-2    & cyan & 2.8 $\pm$ 0.3 & 10.0 $\pm$ 1.1\\
  1a-3     & purple & 2.9 $\pm$ 0.3 & 9.8 $\pm$ 2.8\\
  1a-4       & olive & 2.8 $\pm$ 0.2 & 12.6 $\pm$ 1.1 \\
  NGC2067       & red & 2.8 $\pm$ 0.3 &  1.2 $\pm$ 0.4\\
         {\bf Orion relic filament} & black (arrows) & 2.4 $\pm$ 0.3 & 16.8 $\pm$ 2.5 
    \end{tabular}
    \caption{List of representative groups in the OSFR, as shown in Fig.~\ref{fig:Orion_onsky} in the 
    indicated colour, with their mean parallax and mean age, and their associated range. We link each group to the OSFR division by \protect\cite{Bally2008_book} (the first column of this table). In case we found several sub-structures to one of Bally's group, we indicate this.}
    \label{tab:my_label}
\end{table}

\begin{figure*}
 \scalebox{.5}{\includegraphics{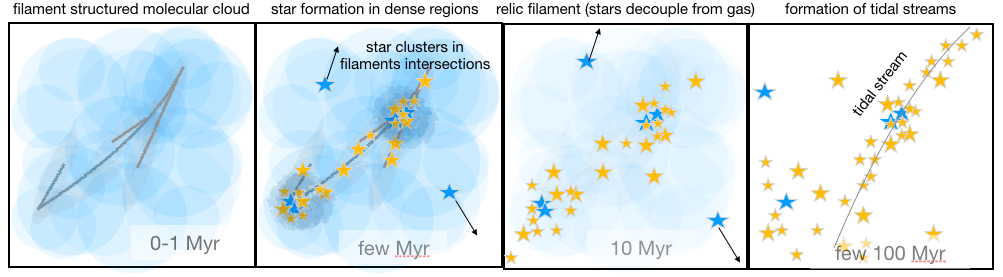}}
 \caption{Sketch of filament star formation. From left to right we show different evolutionary stages starting from a molecular cloud with filament substructure. Stars then form in the densest parts (in the filaments, but mainly in the intersection of sub-filaments of fibres which form embedded clusters). After few Myr the stars decouple from the gas as the residual gas disperses. Some of the embedded clusters will survive, some dissolve after the dispersal of the residual gas. This evolutionary stage is the \textit{relic filament}, i.e., the original molecular cloud filament remains evident in the freshly hatched stellar population. After a couple of 100 Myr the surviving star clusters can start to develop tidal tales. At later stages and after one to a few Galactic orbits (0.2-few Gyr) the relic filament disperses into the Galactic field due to tides and shears.  
}
\label{fig:fil_for}
\end{figure*}

\section{Physical origin of the Orion Relic Filament}

The Orion relic filament is an approximately 90 pc-long structure that is coeval, with an age of $\approx17$Myr. To understand the physical origin of such a structure, its absolute age plays a critical role and can help us to distinguish between different possible formation scenarios. 

A natural scenario for the formation of such a coeval and elongated structure would be that it originated in one star cluster, that was disrupted by removal of the residual gas and further extended by Galactic tides up to the present day size \citep[see discussion in][]{Kroupa2001T,Kroupa2005}.  This is clearly a viable scenario for formation of stellar streams. However, this process acts on a time scale proportional to the Galactic rotation period (200 Myr), and thus cannot be responsible for the formation of the discovered structure. 

In principle, gas removal could trigger the expansion of  a star cluster. For a 17 Myr-old structure spread over 100 pc, this would imply an expansion velocity of approximately $5-10$~km/s. We do not find evidence of such an expansion. 
Moreover, the dispersion of the stellar motions are $\approx 0.5$ km/s and hence too low  (by an order of magnitude) with respect to what would be needed to form such elongated structure. In addition, considering the low number of observed stars, the likely initial stellar mass of the putative star cluster is small ($<< $ 1000 $M_{\odot}$). For such star clusters, the expected expansion velocities are  $\approx$ 0.5 km/s \citep[e.g.][]{Brinkmann2017}, as also recently confirmed by Gaia DR2 data \citep{Kuhn2019a,Kuhn2019b}.

Another possible explanation is that the Orion relic filament is the relic of star formation in a molecular cloud filament structure. We show in Fig~\ref{fig:fil_for} an explanatory sketch of the possible formation scenario. As emphasised by \cite{Andre2014} it is generally accepted that stars and more specifically pre-stellar cores form in regions of the molecular clouds characterized by high mass density (first two panels from the left of Fig.~\ref{fig:fil_for}). \cite{Andre2014} demonstrate that, together with pre-stellar cores, filaments play a critical role in the star formation process as star formation proceeds in them. Moreover, filament intersections can provide column densities and pressures \citep{Myers2009,Myers2011} high enough to trigger the formation of massive stars and embedded star sub-clusters~\citep{Joncour2018, Hill2011, Hennemann2012, Schneider2012}. 

Based on high resolution observations of nearby star-forming regions (Orion, NGC~1333) obtained with the Atacama Large Millimeter/submillimeter Array (ALMA), \citet{Hacar+17,Hacar+17b} provide new strong observational  evidence
in support of a scenario in which pre-stellar cores forms in high-density filaments while embedded star clusters and high-mass stars reside in the intersection of filaments, hubs and ridges~\citep[][]{Hacar2018}. Such filaments can be as large as a few hundred pc in size~\citep{Li2016,Mattern2018,Gorssschedl2018}. 
Moreover, ALMA observations of the merging Antennae galaxies indicate the detection of proto-globular clusters (GC) and suggest that the extreme values of pressure detected in the observed regions may be produced by ram pressure from the collision of filaments \citep{Finn2019}.

The OSRF is a complex star forming region where multiple burst of star formation have been happening over the last 10 to 1{\bf 2} Myr~\citep{Bally2008_book}. While Orion A is a 90pc-{\bf large} star forming region~\citep{Gorssschedl2018}, filamenary structures with ongoing star formations in 
the vicinity of the Trapezium region have been recently observed with ALMA by~\citet[][]{Hacar+17}. In this context, the 17~Myr coeval filament of stars found in this paper can be interpreted as the relic filamentary structure of one of (in not the) oldest start formation event in OSFR.   
In other words, the Orion relic filament might  bear witness to
a filament-driven star formation event that occurred approximately 17~Myr ago in the OSRF and is decoupled from its natal gaseous filament. 
As shown by~\citet[][]{Padoan2017}, the gaseous filaments are expected to survive $\approx$ 10~Myr \citep[see also][]{Egusa2004,Egusa2009,Fukui2010,Meidt2015,Padoan2016} while the stars have lost association with their birth clouds after a few Myr~\citep[][]{Grasha2019}. This scenario is described in the third panel from the left of Fig.~\ref{fig:fil_for}. 

Using Hipparcos, \cite{Bouy2015} report the discovery of a number of large-scale ($\approx$ 100 pc in length) stellar streams in the Solar neighbourhood, Scorpius-Centaurus, Vela and Orion. Their detection was later confirmed by \cite{Zari2018} using Gaia DR2 data. Given the definition above, we suppose that these streams might be too young to be tidal streams and thus are most likely the same kind of objects as the Orion relic filament. 

Strong efforts have been made~\citep[e.g.,][]{Federrath2016,Semadeni2017} to perform hydrodynamical simulations capable of reproducing the collapse of the gas into a sequence of filaments of initially turbulent self-gravitating molecular clouds.  While such simulations
provide theoretical insights into filamentary star formation, they are limited
by the lack of spatial resolution needed to explore the physics of star formation
in filamentary structures at the sub-parsec scale.

Our observations provide unique observational evidence that filamentary star formation is producing large scale co-eval structures, that were likely sub-clustered in the past, but did not merge. 
Hence, observations of relic filaments present a very valuable asset to our understanding of star formation in general. 

\section{Conclusions}

We report the discovery of the oldest stellar population linked to the Orion star forming region. The structure is clustered in proper motions, is located at a distance of $\approx430$ pc from us, has a filament-like shape on the sky of length of $\approx 90$~pc and is coeval with an age of $\approx 17$ Myr. We measure radial velocities (RV) of 13 proper motion-selected candidate members using the high-resolution spectrograph Hermes at the 1.2m Mercator telescope at La Palma. The distribution of the RVs, complemented with a few measurements available from Gaia DR2, peaks at $26.1 \pm 4.7$~km/s, indicating that the group of stars share a common bulk motion. 

 We interpret the discovered the 17 Myr-old structure, which we refer to as the \emph{Orion relic filament}, to be a relic of star formation in a molecular cloud filament. This may represent a short-lived evolutionary phase after gas removal and prior to the dispersion of the structure due to shears and tides.
 The formation of tidal tails and streams should occur later, if at all,  on a time-scale of hundreds of Myr and only for the surviving (i.e., massive and compact enough) star clusters. The following evolutionary stages thus seem to occur:
i) dust/gas filaments and fibres, ii) filament star formation emerging in the densest parts of the molecular cloud filaments, iii) phase of relic filament after the stars decouple from the gas, iv) tidal/shear disruptions and tidal streams development. We present a descriptive sketch in Fig.~\ref{fig:fil_for} as a summary.  

Therefore this discovery puts additional constraints on filamentary star formation that happens on large scales and should be systematically searched for and quantified. Any simulations describing star formation should aim to be able to reproduce these structures that are lacking evidence of hierarchical large-scale merging. This suggest that molecular clouds are not initially gravitationally bound, but produce large scale structures built from filaments that host nests of stars and star clusters at their intersections \citep[e.g.,][]{Joncour2018}.  

\section*{Acknowledgements}
We thank Laurent Eyer and Jiri Zak for carrying out the {\it Hermes} observations as well as the Mercator team for providing support.

This research is based on observations made with the Mercator Telescope, operated on the island of La Palma by the Flemish Community, at the Spanish Observatorio del Roque de los Muchachos of the Instituto de Astrof\'isica de Canarias. {\it Hermes} is supported by the Fund for Scientific Research of Flanders (FWO), Belgium, the Research Council of K.U. Leuven, Belgium, the Fonds National de la Recherche Scientifique (F.R.S.-FNRS), Belgium, the Royal Observatory of Belgium, the Observatoire de Gene\`eve, Switzerland, and the Thu\"uringer Landessternwarte, Tautenburg, Germany. This research was made possible through the use of the AAVSO Photometric All-Sky Survey (APASS), funded by the Robert Martin Ayers Sciences Fund. C.F.M. acknowledges the ESA Research Fellowship. This work has made use of data from the European Space Agency (ESA) mission
{\it Gaia} (\url{https://www.cosmos.esa.int/gaia}), processed by the {\it Gaia}
Data Processing and Analysis Consortium (DPAC,
\url{https://www.cosmos.esa.int/web/gaia/dpac/consortium}). Funding for the DPAC
has been provided by national institutions, in particular the institutions
participating in the {\it Gaia} Multilateral Agreement.
This research has made use of the SIMBAD database,
operated at CDS, Strasbourg, France.

\clearpage
\onecolumn
\bibliographystyle{mnras}
\bibliography{library}
\twocolumn 
\clearpage
\appendix

\appendix

\section{Extra material}

\begin{figure*}
 \scalebox{.8}{\includegraphics{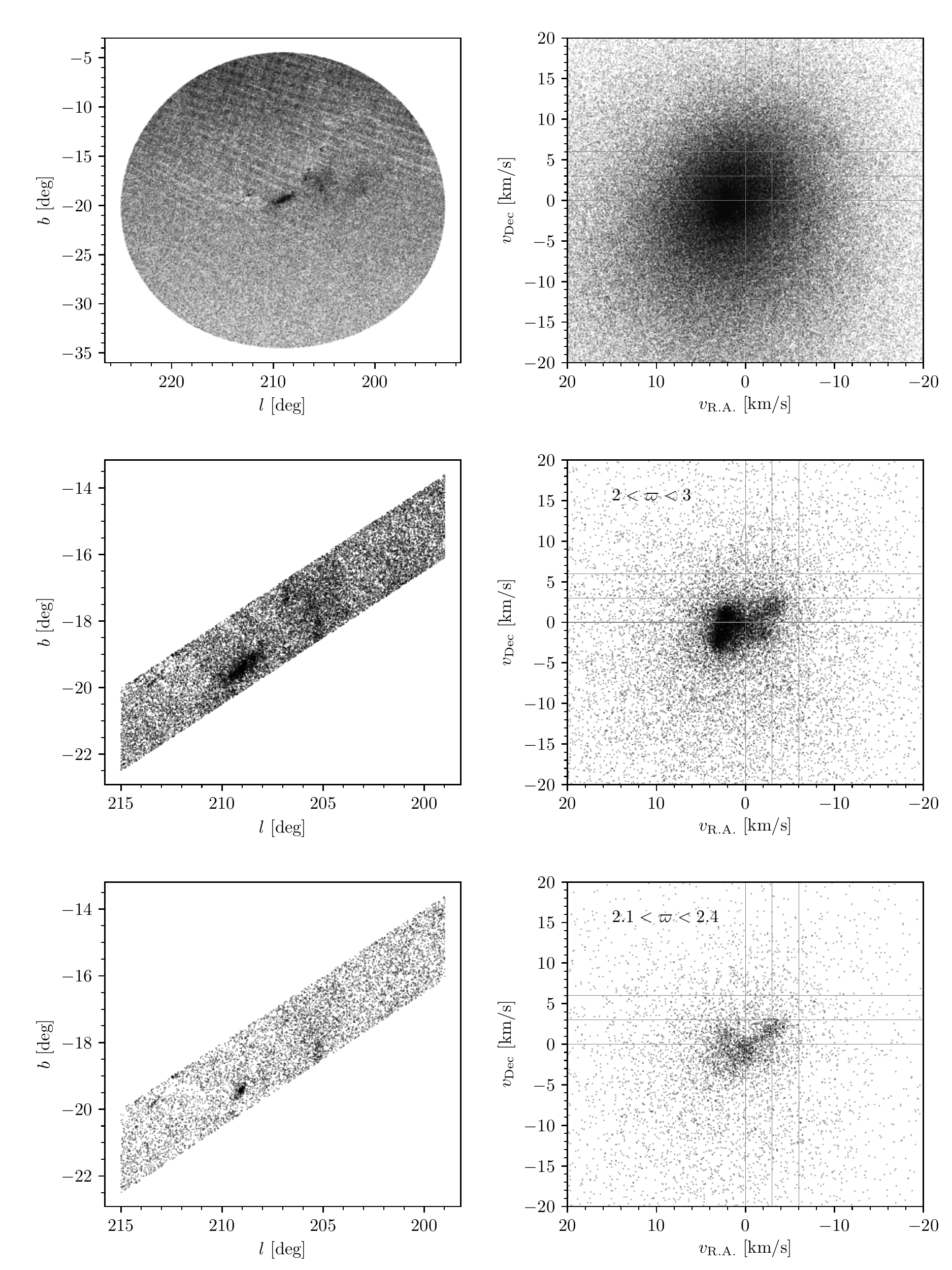}}
 \caption{{\bf Top panels:}  (left) The 15deg catalogue  in $l$ and $b$ coordinates and (right) in proper motion space (in km/s) in R.A. and Dec. {\bf Middle panels:}  (left) Same as above with parallax cut between 2 to 3 mas and the cut in $l$ and $b$ applied. (right) Proper motions corresponding to middle left panel data selection. {\bf Bottom panels:}  Same as the middle panels above, albeit with a stricter cuts on parallax, showing only objects with parallax between 2.1 and 2.4 mas.
}
\label{fig:lb_cut}
\end{figure*}

\end{document}